\documentclass[pre,twocolumn,amssymb,showkeys,superscriptaddress,notitlepage]{revtex4-1}
\pdfoutput=1
\usepackage{graphicx}
\usepackage{color}
\usepackage{epsfig}
\usepackage{latexsym}
\usepackage{bm}
\usepackage{ulem}
\usepackage{color}
\usepackage{enumerate}
\usepackage{enumitem}

\usepackage{dcolumn}
\usepackage{amsmath,amssymb}    
\usepackage{bm} 
\usepackage{hyperref}
\usepackage{latexsym}
\usepackage{color}
\usepackage{subfigure}
\usepackage{fancyhdr}
\usepackage{chngcntr}

\def\beq{\begin{equation}}
\def\eeq{\end{equation}}

\def\beq{\begin{equation}}                           
\def\eeq{\end{equation}}                           
\def\bea{\begin{eqnarray}}                           
\def\eea{\end{eqnarray}}        

\begin{document}

\title{Nonquenched rotators ease flocking and memorise it}


\author{Rakesh Das}
\email[]{rakeshd68@yahoo.com}
\affiliation{ S N Bose National Centre for Basic Sciences, Block JD, Sector III, Salt Lake, Kolkata 700106, India} 
\author{Manoranjan Kumar} 
\email[]{manoranjan.kumar@bose.res.in}
\affiliation{ S N Bose National Centre for Basic Sciences, Block JD, Sector III, Salt Lake, Kolkata 700106, India} 
\author{Shradha Mishra}
\email[]{smishra.phy@itbhu.ac.in}
\affiliation{ Department of Physics, Indian Institute of Technology (BHU), Varanasi 221005, India} 



\begin{abstract}
We introduce a minimal model for a two-dimensional polar flock with nonquenched rotators, and show that the rotators make 
the usual macroscopic long-range order of the flock more robust than the clean system. The rotators memorise the flock-information which helps 
in establishing the robustness. Moreover, the memory of the rotators assists in probing the moving flock. We also formulate 
a hydrodynamic framework for the microscopic model that makes our study comprehensive. Using linearised hydrodynamics, 
it is shown that the presence of such nonquenched heterogeneities increases the sound speeds of the flock. The enhanced sound speeds
lead to faster convection of information and consequently the robust ordering in the system. We argue that similar nonquenched 
heterogeneities may be useful in monitoring and controlling large crowds.
\end{abstract}


\maketitle


\section{Introduction}
Collective behaviour of large-scale systems like crowd of pilgrims 
\cite{BaranwalPLos2015,BarnettPLosOne2016,HughesARFM2003,KimVisuCom2015,CurtisBook2013} 
and flock of birds \cite{cavagnaarcmp2014} spanning a few kilometers, or micron-scale population of bacteria 
\cite{dombrowskiprl2004, asemplosone2017} display many common features like phase segregation \cite{cates2015, klamser2018, Bergmann2018}, 
large density fluctuations \cite{supravat2012}, etc.
These features stem from the self-propulsion nature of the individual constituents exhibiting systematic
movement at the cost of their internal energy \cite{revmarchetti2013,revtoner2005,revramaswamy2010,revvicsek2012,revcates2012}.
A comprehensive understanding of these systems have immense utility \cite{popkin2018} in various aspects of society, viz., safety measures of
human crowds \cite{BaranwalPLos2015,BarnettPLosOne2016,HughesARFM2003,KimVisuCom2015,CurtisBook2013}, cell biology \cite{needleman2017, Xi2019}
and drug delivery employing microrobotics \cite{rajesh2009,lu2018}.
Extensive theoretical and experimental studies have indeed developed a primary insight of the underlying physics in the clean systems
\cite{revmarchetti2013,revtoner2005,revramaswamy2010,revvicsek2012,revcates2012,vicsek1995,tonertu1995,tonertu1998,chate2008,schaller2010}.
However, heterogeneity is inevitable in natural systems \cite{reichhardt2017}, e.g., bacteria moving on complex substrates 
\cite{stearns2011,ponisch2019}, human traffic with varieties of obstacles \cite{zuriguel2018,jiang2014}. 
Recently, various heterogeneous systems are studied, and it is shown that usually the collective movement of the flock gets hampered 
by the heterogeneities \cite{chepizhko2013,marchetti2017,quint2015,rakesh2018,tonerprl2018,tonerpre2018,morin2017}. 
Surprisingly, specific type of external agents can also help in flock formation or crowd control \cite{kulkarni2019,yano2018,zuriguel2011,Lin2017}. 
However, a generic framework for the heterogeneous self-propelled system is lacking in the literature which may help 
in understanding varieties of flocking systems.

Study of the clean self-propelled systems using minimal rule-based microscopic models has been successful in predicting and 
explaining many features therein \cite{vicsek1995,gregoire2004}.
Inspired by this, in this article, we propose a minimal model for a heterogeneous
self-propelled system. The heterogeneity agents are modeled as inertial rotators which try to retain their orientations 
and affect the neighbouring flock. However, the flock also gives feedback of its orientational information to the rotators and changes 
their orientations. Our numerical study reveals that an interplay of the feedback mechanism and the inertia of the rotators 
build a {\it correlated} flock which is more robust to fluctuations
than the clean system. Moreover, these rotators memorise the flock orientation that offers a novel mechanism for probing the 
flock statistics. We also provide a hydrodynamic description of this minimal model. The linearised calculations show that 
these heterogeneities effectively increase the convection speed that establishes the robust ordering in the system. 


\section{Model}
We consider a collection of $N_s$ polar self-propelled particles (SPPs) on a two-dimensional (2D) substrate. These particles are 
characterised by their instantaneous positions ${\bm r}_j(t)$ and orientations $\theta_j(t)$. Each SPP tries to orient parallel 
to its neighbours, although it makes errors. The particles move along their updated orientations with a constant speed $v_s$.
The substrate is also populated with $N_r$ randomly-placed nonquenched rotators (NRs). The NRs remain stationary and have their 
own orientations $\phi_j(t)$. The $j$-th NR influences its neighbouring SPPs and tries to reorient those SPPs along $\phi_j$. In turn, 
the flock of the SPPs also tries to reorient the NR along the mean flock-orientation. However, the effect of the SPPs on a NR is 
suppressed by its inertia, as the NR tries to retain its earlier orientation. Therefore, the model is described by the 
following update rules:
\begin{eqnarray}
 \theta_j(t+1) &=& \arg \left[ \sum_{k \in R} e^{i \theta_k(t)} + \mu \sum_{k \in R} e^{i \phi_k(t)} \right] + \eta_{\theta}\psi_{\theta}, \quad\label{updaterule2}\\
 {\bm r}_j(t+1) &=& {\bm r}_j(t) + {\bm v}_j(t+1),  \label{updaterule1}\\
 \phi_j(t+1) &=& \arg \left[ e^{i \phi_j(t)} + \alpha \sum_{k \in R} e^{i \theta_k(t)} \right],\label{updaterule3}
\end{eqnarray}
where $\arg [{\bm z}]$ represents the argument $\Theta$ of ${\bm z}={\mathcal R} e^{i \Theta}$.
The self-propulsion velocity ${\bm v} = v_s\left(\cos\theta(t),\sin\theta(t)\right)$, $R$ represents the interaction radius, and 
summations are considered over all neighbours within $R$. The mutual interactions among the SPPs and the NRs are tuned by the 
strengths $\mu$ and $\alpha$. We choose these two parameters mutually
independent due to the lack of momentum conservation in the dry active systems \cite{revmarchetti2013}. Errors in the 
process of orientation update of the SPPs are incorporated by an uniform additive noise $\psi_\theta$ in the range $[-\pi,\pi]$ 
with zero mean and white correlations. $\eta_\theta \in [0,1]$ represents the noise strength. The rotators are called nonquenched  
as their orientations are changed by feedback of the flock-orientations.

\begin{figure}[t]
\includegraphics[width=\linewidth]{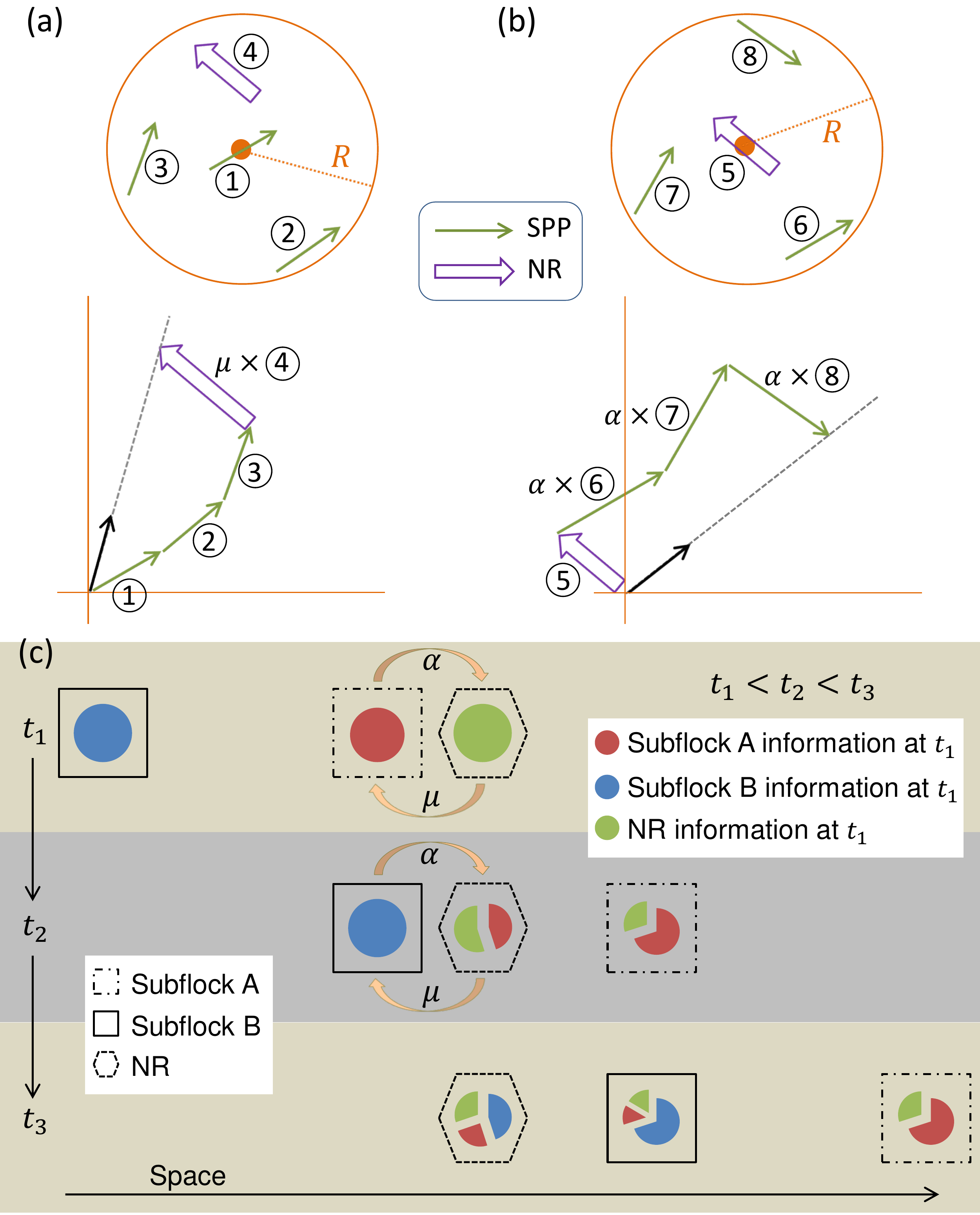}
\caption{
{Schematic of interaction among flocks and NR.}	
	(a) {\it Top} -- A test SPP (`1') interacts with its neighbours (`2, 3, 4') within a distance $R$. 
	{\it Bottom} -- Vectorial representation of the orientation update rule of the SPP (`1'). The SPP-NR (`1'-`4') interaction 
	is tuned by $\mu$. The dashed line shows the resultant orientation and the black arrow on it represents the corresponding 
	unit vector.
	(b) Similar presentation for the orientation update of a test NR (`5'). The NR-SPP interactions are tuned by $\alpha$.
(c) Two subflocks (A -- broken square, B -- solid square) of SPPs and a NR (broken hexagon) are considered in an arbitrary 2D space 
at time $t_1$. The colours inside the geometries indicate informations therein. All the informations are called pure at $t_1$. 
As A and NR are nearby at $t_1$, they interact mutually through $\mu$- and $\alpha$-terms, and meanwhile B remains reclusive. 
During interactions, A and NR exchange informations as per the values of $\mu$ and $\alpha$.
A passes by the NR outside its interaction range by $t_2$. By that time, B comes close to NR and interact mutually. 
Consequently, NR receives information of B, and transfer its current information (which also contains information of A) to B, as 
depicted for a late time $t_3$. Other 
than direct interactions among A and B, this extra means of information transfer through the inertia of NR establishes ordering 
robust than the clean system. Also, NR acts as probe to the flocks as it memorises informations of the flocks.
}
\label{figschematic}
\end{figure}

The effect of the neighbouring particles on a tagged SPP (NR) is schematically presented in Fig.~\ref{figschematic}. For 
an arbitrary configuration shown in the top panel of Fig.~\ref{figschematic}a (respectively, b), the effective interactions are depicted in the bottom 
panel after properly tuned by $\mu$ (respectively, $\alpha$). Due to the interactions with the neighbours, the orientation of the tagged particle would 
be updated towards the resultant direction, as represented by the black arrow. 
We evidence that due to the presence of the NRs, a robust coordination is developed among the SPPs. The corresponding mechanism is depicted in the schematic 
Fig.~\ref{figschematic}c, which we discuss in Sec.~\ref{subsecLRO}. 

The above model is similar to the celebrated Vicsek model \cite{vicsek1995} in the absence of the NRs, and it describes a clean flock
where a true long-range order (LRO) exists in 2D \cite{tonertu1995,tonertu1998}. Also note that the rotators become quenched for $\alpha=0$, and 
therefore, no long-range order but a quasi-long-range order (QLRO) may survive in the system \cite{rakesh2018,tonerprl2018,tonerpre2018}.      

\begin{figure}
\includegraphics[width=\linewidth]{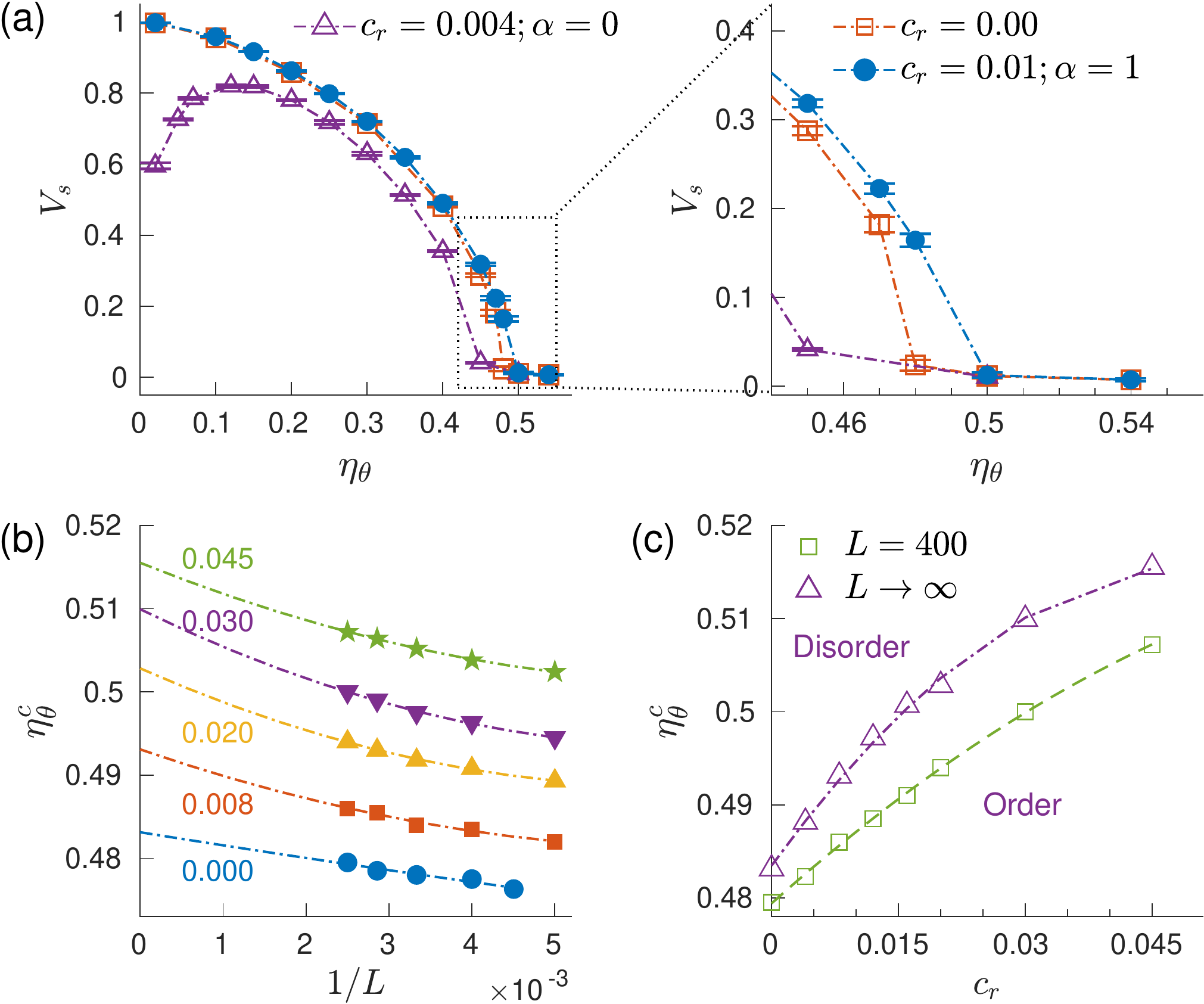}
\caption{
{Robust ordering due to NRs. The data are shown for $\mu=200$.} 
(a) The SPPs show monotonic order-disorder transition with $\eta_{\theta}$ in the presence of the NRs ($c_r=0.01, \alpha=1$),
similar to the clean system ($c_r=0$). Unlike that, the quenched model ($c_r=0.004, \alpha=0$) shows nonmonotonic transition. 
The curves are zoomed near the transition on the right panel of (a).
(b) Critical noise $\eta_{\theta}^{c}$ increases with $L$. The numerics on the left margin indicate $c_r$, and the dashed lines show 	
respective quadratic fits. The thermodynamic-limit value $\eta_{\theta}^c(L\rightarrow\infty)$ of the critical noise is obtained from the 
extrapolation.
(c) $\eta_{\theta}^c(L)$ increases quadratically with density $c_r$. The broken lines show respective fits.
Order and disorder states of the SPPs are indicated for the thermodynamic limit.
}
\label{fig1}
\end{figure}

\section{Results}
\subsection{ Robust long-range order} \label{subsecLRO}
We simulate the update Eqs.~(\ref{updaterule2})-(\ref{updaterule3}) numerically in a $L\times L$ ($L\in[100,500]$) substrate 
with periodic boundaries. $N_s$ SPPs (density $c_s= N_s/L^2 =1$) and $N_r$ NRs (density $c_r = N_r/L^2$) with random orientations 
are distributed on the substrate. As we are interested in the physical limit where heterogeneities are small in numbers as compared 
to the SPPs, we restraint ourselves to $c_r\in[0,0.045]$. Further, we consider $v_s=1$ and $R=1$, and execute a code exploiting OpenMP 
application program interface. We note that depending on the system size, $3\times10^5$ to $16\times10^5$ iterations are sufficient to 
attain steady states, and calculate the relevant quantities by averaging over next $5\times 10^5$ to $14\times 10^5$ iterations. Upto 
$30$ realisations are used for better statistics. The order parameter of the SPPs 
$V_s= \langle V_s(t)\rangle_t = \langle \frac{1}{N_s}\left|\sum_{j=1}^{N_s}e^{i\theta_j(t)} \right|\rangle_t$ 
varies from zero to unity for a disordered to an ordered state, respectively.

The clean system ($c_r=0$) shows a monotonic order-disorder transition with increasing $\eta_{\theta}$ (Fig.~\ref{fig1}a). However, 
in the presence of the quenched rotators ($\alpha=0$), the system achieves optimal ordering at a nonzero $\eta_{\theta}$. This optimality 
feature emerges as the quenched rotators disturb transfer of informations among the SPPs, and the system needs a certain noise to 
circumvent that hindrance \cite{chepizhko2013,rakesh2018}. As the optimal $\eta_\theta$ increases with $\mu$, we set $\mu=200$ such 
that the optimal $\eta_\theta$ attain moderate values for the studied range of $c_r$. In contrast to the quenched model, the system 
shows a monotonic transition for $\alpha>0$ in the presence of the NRs. 
Similar to the clean case \cite{vicsek1995,chate2008}, we note a homogeneous ordered state of SPPs in the presence of the NRs for noises much 
smaller than its critical value for the order-disorder transition. We also note that banded configuration emerges near the critical 
noise where a highly ordered dense cluster of SPPs travels over disordered sparse background \cite{chate2008,aldana2009}.
Surprisingly, the system with 
the NRs survives upto $\eta_{\theta}$ higher than its clean counterpart, as the zoomed version of Fig.~\ref{fig1}a shows. This 
necessarily implies that the flock is more robust to the external fluctuations ($\eta_\theta$) as compared to the clean system. To 
ensure this emerged robustness, we calculate the critical noise $\eta_\theta^c$ of the order-disorder transition and compare it for 
several $c_r$. The Binder cumulant \cite{binder1981,chate2008,aldana2009} 
$U(L) = 1 - \langle V_{s}^4(t,L) \rangle_t / 3\langle V_{s}^2(t,L) \rangle_t$ 
shows a dip to negative values near the transition (Appendix~\ref{appendix_sec1}). We define $\eta_\theta^c$ as the noise corresponding 
the $\min(U(L))$. 
We check that reduced critical noise $\epsilon = 1 - \eta_\theta^c / \eta_\theta^{c {\mathcal L}}$ scales as $L^{-\zeta}$, where 
$\eta_\theta^{c {\mathcal L}}$ represents critical noise obtained for $L=400$ (data not shown). We note that the finite size scaling (FSS) 
exponent $\zeta=2$ for the clean system, as reported in Ref.~\cite{chate2008}, and the exponent increases for $c_r>0$. Though it is interesting 
to check the variation in $\zeta$ with $c_r$, a rigorous FSS study of the present Vicsek-like model with {\it angular} noise \cite{chate2008} 
is numerically expensive and beyond our current objective. Rather, we note that $\eta_\theta^c$ increases quadratically with decreasing $1/L$ (Fig.~\ref{fig1}b). 
So, the thermodynamic limit values 
$\eta_\theta^c(L\rightarrow\infty)$ of the critical noises are obtained from extrapolations of these curves to the $L\rightarrow\infty$ limit. 
Interestingly, $\eta_{\theta}^c(L\rightarrow\infty)$ increases quadratically with $c_r$ (Fig.~\ref{fig1}c), and therefore, the NRs indeed 
offer a mechanism for a flock to be more robust than the clean system. 
We also plot the $\eta_\theta^c$-$c_r$ curve for $L=400$ for comparison, and note similar behaviour as explained for the thermodynamic limit.

\begin{figure}[b]
\includegraphics[width=\linewidth]{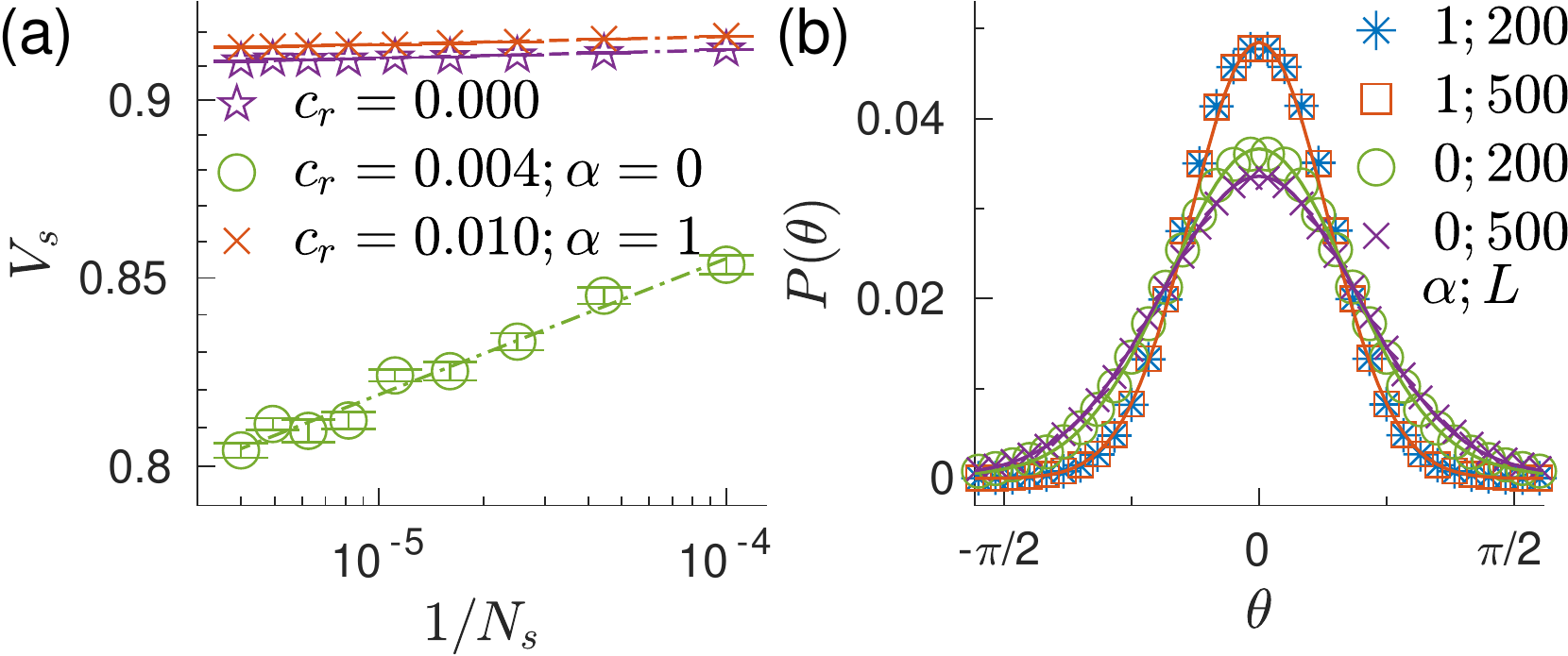}
\caption{
LRO in the presence of NRs. The data are shown for $\mu=200$ and $\eta_\theta=0.15$. 
(a) The SPP-order parameter does not vary with system size for the clean system ($c_r=0$) or the system with the NRs ($c_r=0.01, \alpha=1$). 
However, in the presence of the quenched rotators ($c_r=0.004, \alpha=0$), the order parameter decreases algebraically. The lines 
show respective algebraic fits. 
(b) The orientation distribution of the SPPs do not depend on the system size in the presence of the NRs, whereas the distribution 
broadens with $L$ for the quenched model. 
}
\label{fig2}
\end{figure}

The order parameter of the clean system does not depend on the system size (Fig.~\ref{fig2}a) which is a manifestation of the
LRO therein \cite{vicsek1995, tonertu1995, tonertu1998, rakesh2018}. We also note that in the presence of the quenched rotators, 
$V_s$ decreases algebraically with $N_s$, implying a QLRO state in the system \cite{rakesh2018,tonerprl2018,tonerpre2018}. However, 
$V_s$ does not change with $N_s$ in the presence of the NRs. Moreover, the order parameter in the presence of the NRs is larger than 
clean system. Therefore, similar to the clean system, a LRO exists in the presence of the NRs, and that state is more robust 
than the clean system. The nature of the ordered state is further confirmed by calculating a normalized distribution $P(\theta)$ of 
the SPP-orientations for various system sizes \cite{rakesh2018}. This distribution is a measure of the orientation fluctuations among 
the SPPs, and it does not vary with the system size in the presence of the NRs (Fig.~\ref{fig2}b). However, $P(\theta)$ widens with 
system size in the quenched model as there exists a QLRO only \cite{rakesh2018}. 

The mechanism that makes the LRO state more robust than the clean system can be understood from the schematic Fig.~\ref{figschematic}c drawn on an 
arbitrary 2D space. Two subflocks A and B, and a NR have their {\it pure} orientational informations at time $t_1$, as represented by the colours red, 
blue and green, respectively. While B is away from the NR, A and NR interact mutually through the $\mu$- and $\alpha$-terms and exchange informations. 
Therefore, at a later time 
$t_2$, A and  NR contain both the pure informations (red and green) corresponding to time $t_1$. The proportion of this exchange 
is determined by the parameters $\mu$ and $\alpha$. However, by that time A moves beyond the interaction range of the NR, and B 
comes close to the same NR. Due to a similar kind of interaction through $\mu$ and $\alpha$, the NR now receives the pure information 
of B (blue). In turn, B receives the pure information of A (red) even without any direct interaction, as shown in the schematic 
for time $t_3$. Therefore, in the duration of $t_1$ to $t_3$, the NR has received feedback from the subflock A and later it 
delivered that information to B through its inertia. In addition to the usual convection of the SPPs \cite{vicsek1995,tonertu1998}, 
this mechanism provides an extra means of information transfer among the particles that induces robustness in the system. 
Also note that the NR memorises the informations of the passing by flocks, which is beneficiary for probing the flocks through
these external agents, as we discuss next.


\subsection{NRs probe the flock}
We investigate the orientation autocorrelation of the NRs, defined as
$C_{\phi}(t) = \langle \cos\left[\phi_j\left(t+t_0\right)-\phi_j\left(t_0\right)\right] \rangle_{j,t_0}$. 
Here $\langle\cdot\rangle$ symbolizes averaging over all the NRs and many steady-state reference times $t_0$. Starting from an 
orientation $\phi_j\left(t_0\right)$, $j$-th NR changes its orientation due to the feedback from the flocks. Therefore, $C_\phi$ 
shows an early-time decay, and beyond that it saturates to the square of the order parameter $V_r$ of the NRs. $V_r$ is defined 
similar to $V_s$. The early-time decay in $C_\phi$ indicates the timescale upto which a NR remembers its earlier orientation. 
This timescale is necessarily dictated by $\alpha$, as Eq.~(\ref{updaterule3}) suggests. $V_r$ depends on $\eta_\theta$ 
through the feedback from the flock, and also on $\alpha$. However, $V_s$ does not change significantly with $\alpha$, provided $\alpha>0$. 
Therefore, we stress that the flock-phenomenology described here and its implications hold for any finite value of $\alpha$. 
The results presented in this article are obtained for $\alpha=1$.

\begin{figure}[t]
\includegraphics[width=\linewidth]{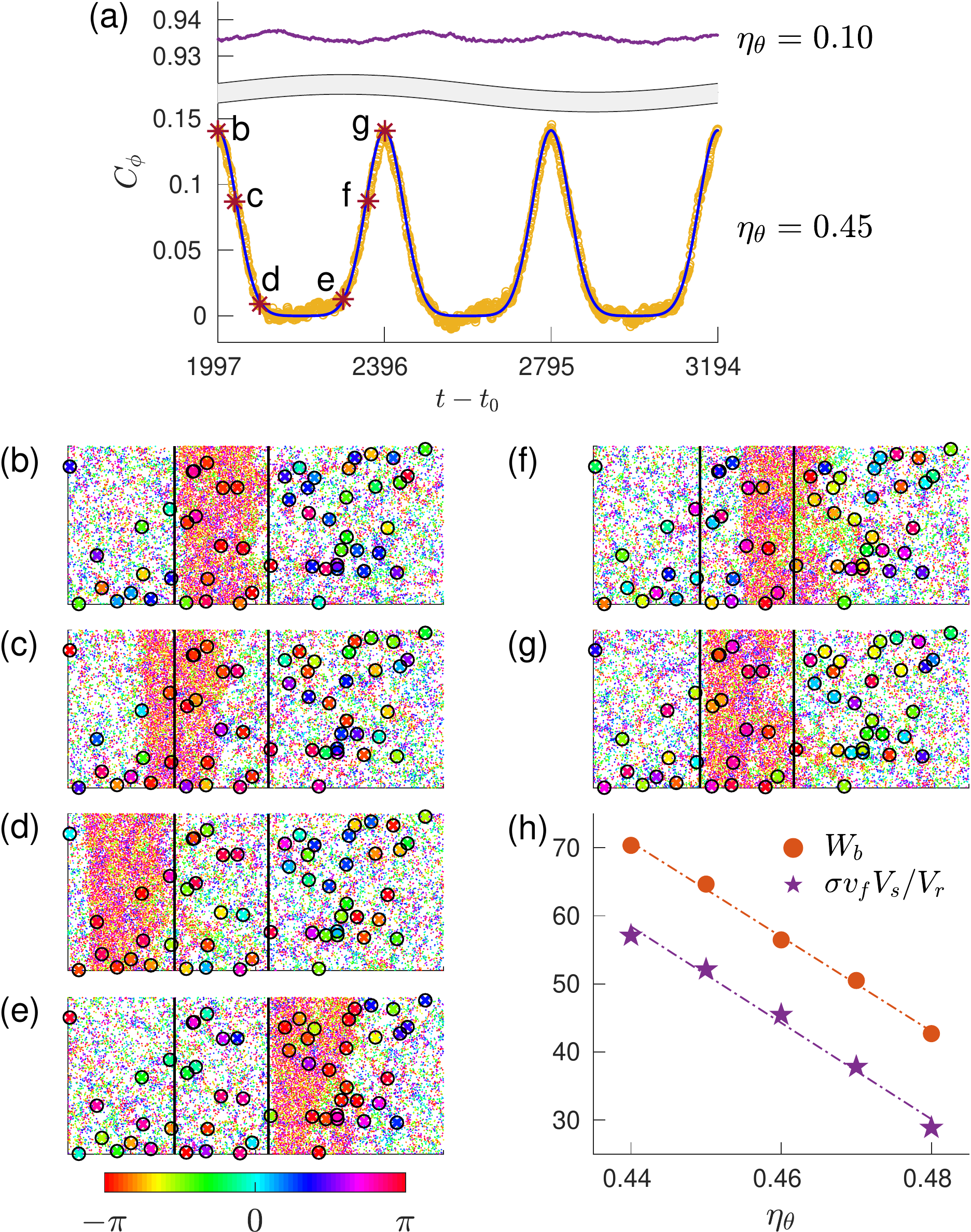}
\caption{
{Flock information is stored in the NRs. The data are shown for $\mu=200$.} 
(a) Autocorrelation function $C_\phi$ of the NRs ($c_r=0.01, \alpha=1, L=300$) are shown at the late time, $t_0$ being the reference steady-state 
time. $C_\phi$ shows modulations over its saturated value $V_r^2$, and these modulations are prominent near the critical point ($\eta_\theta=0.45$) 
where band emerges. The solid line shows Gaussian fit to these modulations, and the labels correspond to the snapshots shown in (b-g). 
The snapshots are shown for a zoomed substrate of dimension $300\times100$ for clarity. The black circles indicate the NRs (only few 
shown), and the colours represent orientations of the SPPs and the NRs. Separation in the black vertical lines in (b) gives an estimate of 
the bandwidth $W_b$. $C_\phi$ assumes a finite value only if the current position of the band overlaps to that in (b). 
(h) $W_b$ and $\sigma v_f V_s/V_r$ decreases with $\eta_\theta$ with the same slope. Here the flock speed $v_f=L/\tau$ where 
$\tau$ is the peak-to-peak separation of $C_\phi$ modulations.
}
\label{fig3}
\end{figure}

We note that $C_{\phi}$ shows periodic modulations over its saturated value (Fig.~\ref{fig3}a). This is more prominent near the 
order-disorder transition where band emerges in the system (Fig.~\ref{fig3}b-g) \cite{chate2008,solon2013,solon2015}.   
The modulations in $C_\phi$ offer a novel mechanism of probing the statistics of the flock. Let us consider a banded-state 
configuration at a reference time $t_0$ when all the NRs have their initial orientations $\phi_j\left(t_0\right)$ (Fig.~\ref{fig3}b). 
The NRs inside the band have a spatially correlated distribution of orientations due to the feedback from the correlated SPPs.
However, the NRs outside the band have random $\phi_j\left(t_0\right)$. Though $\phi_j$ changes as the band moves forward, the same 
NRs that are still inside the band (Fig.~\ref{fig3}c) remains correlated to their initial orientations. The rest of the NRs yield a zero 
contribution to $C_\phi$ as they have been averaged over random terms. Therefore, we obtain a finite $C_\phi$ in aggregate. The band 
moves further in the course of time (Fig.~\ref{fig3}d-e), and there exists no finite contribution to $C_{\phi}$ until the band-front reaches 
(due to periodic boundary condition) the region of the band-tail of time $t_0$ (Fig.~\ref{fig3}e). $C_\phi$ again becomes finite with 
time (Fig.~\ref{fig3}f), and it is the maximum when the band reaches the same position as $t_0$ (Fig.~\ref{fig3}g). Following this, 
the same dynamics continues. Therefore, the peak-to-peak separation $\tau$ in $C_{\phi}$ modulations (Fig.~\ref{fig3}b) is the time 
required by the flock (density wave) to traverse the system once (Supplementary Movie \cite{supplemov}). We obtain the flock (band) speed 
$v_f$ from the ratio of $L$ and $\tau$.  The modulations in $C_{\phi}$ shows good fit with Gaussian curves (Fig.~\ref{fig3}a). 
Standard deviation $\sigma$ of the Gaussian fit multiplied by $v_f$ shows a linear decay with $\eta_{\theta}$ (Fig.~\ref{fig3}h). 
Note that $\sigma v_f$ is a NR-property that should give a measure of the width of the band, as the finite contributions to $C_\phi$ 
are due to the band. Next, we verify that $\sigma v_f$ indeed bears the informations of the band.

We divide the whole substrate into $L$ strips of unit width parallel to the band. The ratio of the number of SPPs inside a strip to 
$L$ gives the local density $c_s^{loc}$ therein. The lateral dimension for which $c_s^{loc} > c_s$ is defined as the bandwidth, which 
is similar to the length between the black vertical lines in Fig.~\ref{fig3}b. We calculate the mean bandwidth $W_b$ averaged 
over many snapshots and see that $W_b$ decreases linearly with $\eta_{\theta}$ with the same slope as  $\sigma v_f V_s/V_r$ versus 
$\eta_{\theta}$ curve (Fig.~\ref{fig3}h). 
Therefore, $\sigma v_f$ which is a NR-property indeed bears the flock information, and the NRs act as footprints of the passed-by flock. 
Note that $W_b$ is calculated considering clustering of the SPPs, whereas $\sigma v_f$ contains information of the ordering in the system. 
Therefore, we need the multiplication factor -- relative strength of order of the SPPs and the NRs ($V_s/V_r$) to compare 
a NR-property $\sigma v_f$ with a flock-property $W_b$.  

Note that the bands may arise in any direction. However, we present the case of the lateral bands only, and emphasize that the same 
argument holds for other band-directions. Also note that similar modulations in $C_\phi$ are present in any non-banded ordered 
state, as shown for $\eta_\theta=0.10$ in Fig.~\ref{fig3}a. However, we present our argument for the banded state only as band  
provides a precise measure of the relevant length and time scales of the flock dynamics. Therefore, the flock informations can be 
obtained from the modulations in $C_\phi$ of the NRs, where the amplitude and the frequency of the modulations are set by the 
ordering and the extent of the flock.


\subsection{Hydrodynamic description}
We develop a hydrodynamic framework for the nonquenched model described above. The relevant 
slow fields of the system with the NRs are -- (i) density $\rho({\bm r},t)$ of the SPPs, (ii) polarisation ${\bm P}({\bm r},t)$ of 
the SPPs and (iii) polarisation ${\bm P_r}({\bm r},t)$ of the NRs. The density $\rho_r$ of the NRs is considered uniform 
as they are immobile and randomly distributed. Following the phenomenology of the system, we write the hydrodynamic equations of 
motion (EOMs) for the slow variables as follows: 
\begin{eqnarray}
 \partial_t\rho &=& -v_s\nabla\cdot\left(\rho{\bm P}\right) + D_{\rho}\nabla^2\rho, \label{EOM1} \\
 \partial_t{\bm P} &=& \left\{\alpha_1(\rho)-\beta_1\left|{\bm P}+{\bm P_r}\right|^2\right\}{\bm P} + \lambda_1\left({\bm P}\cdot\nabla\right){\bm P} \notag \\ 
               && \qquad      - \frac{v_s}{2\rho}\nabla\rho + D\nabla^2{\bm P} + \gamma_1\rho{\bm P_r}, \label{EOM2} \\
 \partial_t{\bm P_r} &=& \gamma_2\rho_r{\bm P} - \beta_2\left|{\bm P}+{\bm P_r}\right|^2{\bm P_r}. \label{EOM3}
\end{eqnarray}
The density $\rho$ of the SPPs being a globally conserved quantity, Eq.~(\ref{EOM1}) represents a continuity equation, however with an 
active current contribution \cite{tonertu1998}. Here $v_s$ represents self-propulsion speed, and $D_{\rho}$ is the diffusion coefficient. 
As the ordering of the NRs affects the the SPPs, the mean-field term within the curly brackets in Eq.~(\ref{EOM2}) contains 
both ${\bm P}$ and ${\bm P_r}$. We have considered only the $\lambda_1$ convective term as this is the most relevant convective 
nonlinearity \cite{tonertu1998,rakesh2018,tonerprl2018,tonerpre2018}. The coefficient of the pressure term due to fluctuation 
in $\rho$ is taken $v_s$ for simplicity. Eq.~(\ref{EOM2}) is written under equal-elastic-constant approximation \cite{degennesbook}.
The ${\bm P_r}$ term in Eq.~(\ref{EOM2}) represents the feedback of the NRs to the SPPs that 
indeed depends on the density of the SPPs. Similar to this feedback term, the feedback of the SPPs to the NRs is represented by the 
first term on the right hand side of Eq.~(\ref{EOM3}). The $\beta_2$ term in Eq.~(\ref{EOM3}) stabilizes ${\bm P_r}$ in 
the steady state.


\subsubsection{Mean-field study}
Let us first consider a broken symmetry steady state such that a homogeneous solution of Eqs.~(\ref{EOM1})-(\ref{EOM3}) is 
given by $\rho=\bar\rho, {\bm P}=P\hat\parallel$, and ${\bm P_r}=P_r\hat\parallel$. Here $\hat\parallel$ is the unit vector along 
the broken symmetry direction, and $\hat\perp$ is normal to that. For this homogeneous steady state, we obtain $P_r = BP$, where 
$B=\beta_1\gamma_2\rho_r/\alpha_1(\bar\rho)\beta_2$ (see Appendix~\ref{appendix_sec2}). Using this expression for $P_r$, we get 
\begin{equation}
 P^2 = \frac{\alpha_1(\bar\rho)}{\beta_1} \times \frac{1+\gamma_1\bar\rho B / \alpha_1(\bar\rho)}{(1+B)^2}. \label{eqPsquare}
\end{equation}
Note that $B=0$ for the clean system, and therefore the order parameter is $\mathcal P = \sqrt{\alpha_1(\bar\rho)/\beta_1}$.
Therefore, the order parameter $P$ of the SPPs in the presence of the NRs is greater than its value $\mathcal P$ in the clean system, 
provided $\gamma_1\bar\rho-\beta_1\gamma_2\rho_r/\beta_2 > 2\alpha_1(\bar\rho)$, which indeed holds for an ordered state 
($\alpha_1(\bar\rho)>0$). This necessarily implies a positive shift in the critical point, and therefore validates the existence of a more 
robust ordered state in the presence of the NRs. 


\subsubsection{Linearised hydrodynamics}
Beyond the mean-field calculations, we consider small fluctuations in the slow fields:
\begin{eqnarray}
 \rho = \bar\rho +&& \delta\rho, \qquad {\bm P} = \left(P+\delta P_{\parallel},\delta P_{\perp}\right), \notag \\ 
                              && {\bm P_r} = \left(P_r+\delta P_{r \parallel},\delta P_{r \perp}\right).  \label{eqfluc}
\end{eqnarray}
Simplifying the hydrodynamic EOMs under linearised approximation and solving for the fluctuations $\delta P_\parallel, \delta P_{r \parallel}$ 
and $\delta P_{r\perp}$, we obtain equations for $\delta\rho$ and $\delta P_{\perp}$ as
\begin{eqnarray}
 \partial_t\delta\rho = \left( D_{\parallel}\partial_{\parallel}^2 + D_{\rho}\partial_{\perp}^2 \right)\delta\rho -&& 
                        Xv_sP\partial_{\parallel}\delta\rho - v_s\bar\rho\partial_{\perp}\delta P_{\perp}, \label{eqdelrho} \quad \\
 \partial_t\delta P_{\perp} = D\nabla^2\delta P_{\perp} + \lambda_1 && P \partial_{\parallel}\delta P_{\perp} - 
                              \frac{v_s}{2\bar\rho}\partial_{\perp}\delta\rho, \label{eqdelPperp} 
\end{eqnarray}
where
\begin{eqnarray}
 D_{\parallel} = D_{\rho} + \frac{v_s^2}{2\left(\alpha_1^{\prime}-A^{\prime}\right)}, \quad 
 X = 1 + \frac{\gamma_1B\bar\rho}{\alpha_1^{\prime}-A^{\prime}}. \label{eqDparaX}
\end{eqnarray}
Here $D_\parallel$ is an effective diffusivity, and the factor $X$ tunes the convective speed.
In the clean system, $D_\parallel=D_\rho+v_s^2/4\alpha_1(\bar\rho)={\mathcal D}$ and $X=1={\mathcal X}$. For a finite $\rho_r$, 
$D_\parallel < {\mathcal D}$ and $X>{\mathcal X}$, as shown in Fig.~\ref{fig5}a-b (also see Appendix~\ref{appendix_sec3}). 
Therefore, 
the presence of the NRs reduces the effective diffusivity in the $\parallel$-direction and also increases the convective speed. 
These two modifications in the physical parameters are responsible for faster transfer of informations among the SPPs.

\begin{figure}[t]
\includegraphics[width=\linewidth]{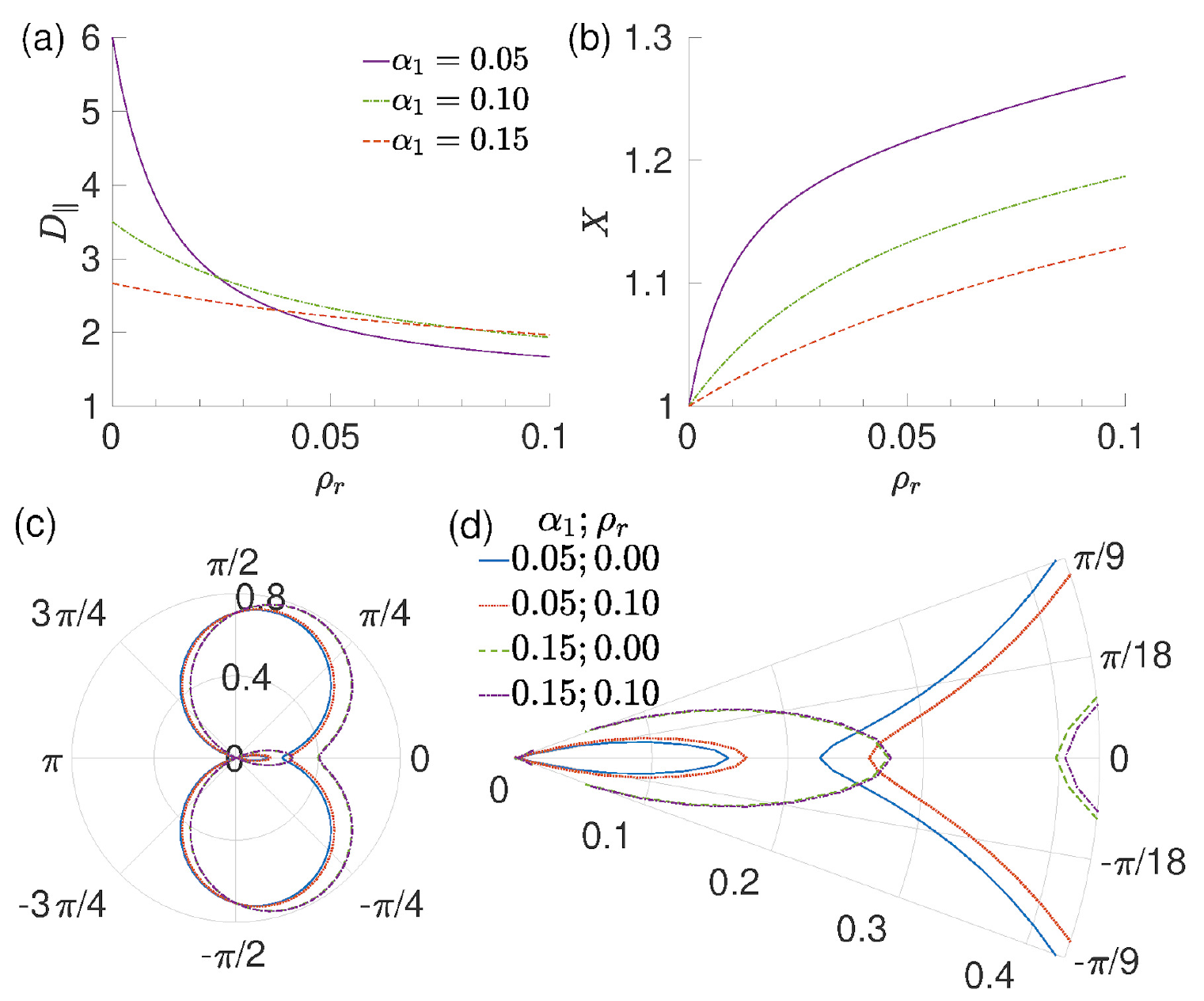}
\caption{
{NRs increase sound-mode speeds.} 
(a) The effective diffusivity $D_\parallel$ decreases and (b) the factor $X$ increases with increasing density $\rho_r$ of 
heterogeneity. The changes are more dominant for smaller $\alpha_1(\bar\rho)$, i.e., near the transition. Legends are the 
same in (a) and (b).
(c) $c_+(\varphi)$ is plotted in polar coordinates for various $\alpha_1({\bar\rho})$ and $\rho_r$, and its zoomed version is 
shown in (d). 
The numerics adjacent to the radial and the angular grids indicate respective values, and help to realize the magnification 
in (d) as compared to (c).
(c) and (d) share the same legends. The mode speed increases with $\rho_r$ along the flock direction, which is 
more prominent for small $\alpha_1(\bar\rho)$.
}
\label{fig5}
\end{figure}

Solving Eqs.~(\ref{eqdelrho}) and (\ref{eqdelPperp}) in Fourier space \cite{tonertu1998,rakesh2018}, we obtain two sound 
modes of the fluctuations as
\begin{eqnarray}
 \omega_{\pm} = c_{\pm}(\varphi)q - i\Gamma_{\rho}\left[\frac{v_{\pm}(\varphi)}{2c_2(\varphi)}\right]
                                  - i\Gamma_{P}\left[\frac{v_{\mp}(\varphi)}{2c_2(\varphi)}\right]. \label{appeq18} 
\end{eqnarray}
Here $\varphi$ is the angle between the wave vector $\bm q$ and the broken-symmetry direction so that 
${\bm q}=\left(q\cos\varphi,q\sin\varphi\right)$, and 
\begin{eqnarray}
 c_2(\varphi) &&= \sqrt{\frac{1}{4}\left(Xv_s+\lambda_1\right)^2P^2\cos^2\varphi + \frac{1}{2}v_s^2\sin^2\varphi}, \qquad \label{appeq19}\\
 c_{\pm}(\varphi) &&= \frac{1}{2}\left(Xv_s-\lambda_1\right)P\cos\varphi \pm c_2(\varphi), \label{appeq20} \\
 v_{\pm}(\varphi) &&= c_2(\varphi) \pm \frac{1}{2}\left(Xv_s-\lambda_1\right)P\cos\varphi. \label{appeq21}
\end{eqnarray}
The wave-vector dependent dampings are 
 $\Gamma_{\rho}({\bm q}) = D_{\parallel}q_{\parallel}^2 + D_{\rho}q_{\perp}$ and
 $\Gamma_P({\bm q}) = Dq^2$.
The sound speeds $c_{\pm}(\varphi)$ differ mutually by a phase shift as $c_+(\varphi) = -c_-(\varphi+\pi)$.
In Fig.~\ref{fig5}c-d we plot $c_+(\varphi)$ in polar coordinates for different values of $\alpha_1(\bar\rho)$. 
For comparison we also plot $c_+$ for the clean system. Clearly, $c_+(\varphi)$ is larger for finite 
$\rho_r$, and the effect is the most dominant for $\varphi=0$. Also note that smaller the value of $\alpha_1(\bar\rho)$, i.e., 
as we approach close to the critical point, more the change in sound speed. Therefore, the effect of the NRs are more prominent 
near the transition, as we have shown earlier in Fig.~\ref{fig1}a. 


\subsection{Fluctuations in the NRs change the scenario}
We also study the effect of an additive noise in the $\phi$-update. Contrary to Eq.~(\ref{updaterule3}), the modified update rule 
for $\phi$ reads 
\begin{eqnarray}
 \phi_j(t+1) &=& \arg \left[ e^{i \phi_j(t)} + \alpha \sum_{k \in R} e^{i \theta_k(t)} \right] + \eta_{\phi}\psi_{\phi}.\label{updaterule3mod}
\end{eqnarray}
The inclusion of this additional noise introduces randomness in the SPP-dynamics through the feedback and inertia mechanism discussed 
above. Consequently, the optimality feature analogous to the quenched model emerges in the system. 
The optimal $\eta_\theta$ for this modified model decreases linearly to zero as $\eta_\phi$ approaches zero. This verifies 
the claim of a monotonic order-disorder transition in the nonquenched model discussed previously in this article. 
We also note that $V_s$ decreases quadratically with $1/N_s$ for this modified model, and an extrapolation of the quadratic fit 
suggests a finite $V_s$ in the thermodynamic limit. Therefore, we stress that the system remains in the LRO state for a 
finite $\eta_{\phi}$. We have discussed this phenomenology in details in Appendix~\ref{appendix_sec4}.


\section{Discussion}
In summary, we study a polar self-propelled system with nonquenched rotators using a minimal rule-based microscopic model, and provide 
a hydrodynamic description of it. Although the self-propelled systems like a collection of bacteria or human crowds differ 
in their specific details 
\cite{BaranwalPLos2015,BarnettPLosOne2016,HughesARFM2003,KimVisuCom2015,CurtisBook2013,cavagnaarcmp2014,dombrowskiprl2004, asemplosone2017},
in general they follow common symmetries and conservation laws \cite{revramaswamy2010}. Usually microscopic models \cite{vicsek1995,gregoire2004} 
are prolific in illustrating the common features of these systems, and hydrodynamic frameworks developed on the basis of the 
microscopic models make the descriptions more general. 

In our model both the self-propelled particles (SPPs) and the rotators feedback each other their orientational informations. As
the rotators memorise the passed-by subflock of SPPs and transfer that information to the next subflock, they offer
an additional means of information transfer. Consequently, these rotators establish a long-range order flock
more robust than the clean system. Note that in the absence of the inertia term in Eq.~(\ref{updaterule3}), a rotator `forgets' 
the information of a passed-by flock immediately, and therefore, the reported robustness vanishes.
The hydrodynamic framework of the nonquenched model verify the enhanced ordering
using mean-field calculations. Further, considering linear fluctuations on a homogeneous ordered state,
we show that the nonquenched heterogeneities decrease the effective diffusivity along flock-orientation
that suppresses the fluctuations. Also, the heterogeneities increase the sound-mode speed which makes the transfer of information 
faster. These general results can be equally applicable to large social gatherings with
similar nonquenched heterogeneities, and tuning the heterogeneities the panic or stampede like
situations can be controlled.

Interestingly, the rotators store the information of the passed-by flock. Therefore, by probing a less number of heterogeneity
agents, we can investigate the flock-statistics, as we have done by comparing the rotator-autocorrelation with the
ordering and extent of the flock. This offers a novel mechanism for monitoring large crowd, alternative to bluetooth- or GPS-based
methods \cite{shah2015,flack2018,gaia2008}.


\begin{acknowledgements}
RD acknowledges CRAY supercomputing facility at S. N. Bose National Centre for Basic Sciences.
SM acknowledges S. N. Bose National Centre for Basic Sciences for kind hospitality during her visit there.
\end{acknowledgements}


\appendix
\counterwithin{figure}{section}


\section{Determining $\eta_\theta^c$ from Binder cumulant} \label{appendix_sec1}
\begin{figure}[b]
\includegraphics[width=0.65\linewidth]{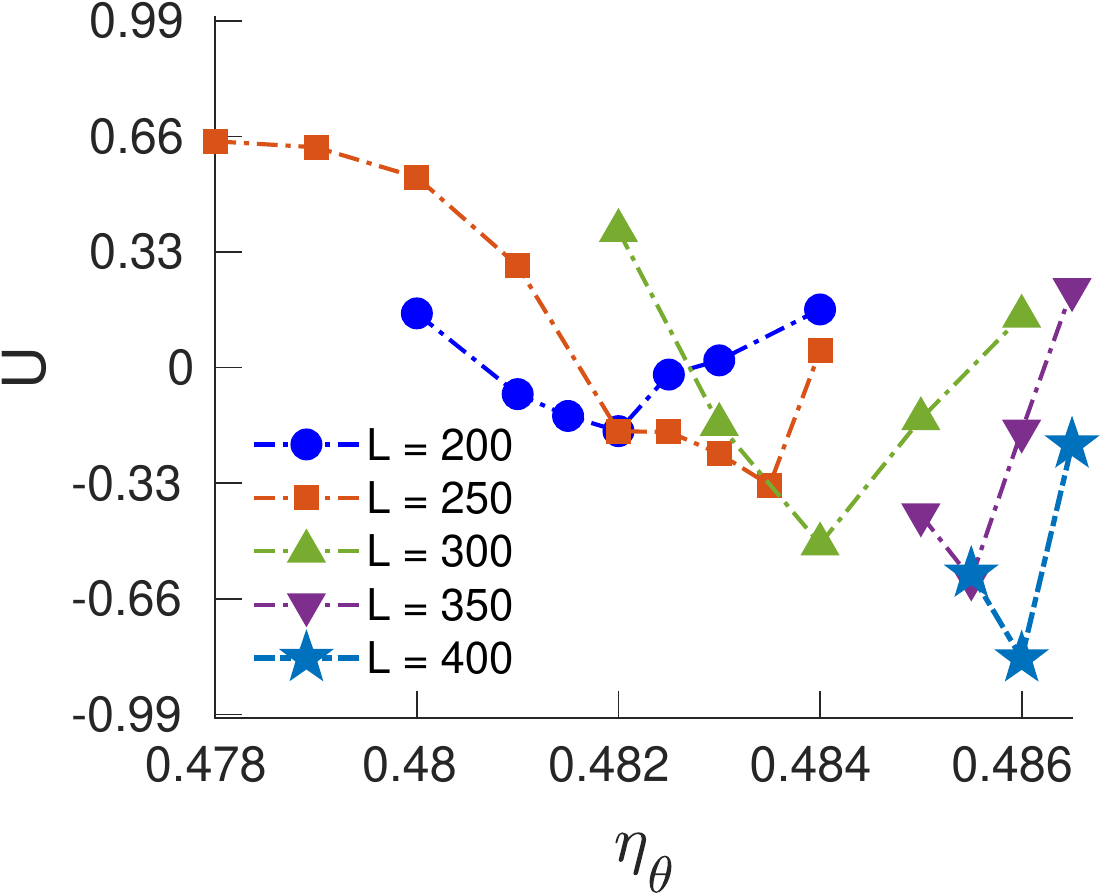}
\caption{
{Critical noise $\eta_\theta^c$ increases with $L$.}
Binder cumulants are shown for $c_r=0.008$, $\mu=200$ and $\alpha=1$, and zoomed near their minima for the sake of clarity. 
}
\label{figbinder}
\end{figure}

We calculate Binder cumulant $U(L)$ for various system sizes. 
$U(L)$ assumes values $2/3$ and $1/3$, respectively, deep in the ordered 
and disordered states, as is expected for a 2D model with continuous rotational symmetry \cite{chate2008,aldana2009}. 
However, it dips to negative values near the transition, as shown in Fig.~\ref{figbinder}, and it indicates the discontinuous 
nature of the transition. We assume the noise corresponding to the minimum of $U(L)$ as $\eta_\theta^c(L)$ which is expected 
to converge to the critical noise \cite{chate2008}. 
Note that $\eta_\theta^c(L)$ increases with $L$, which we fit using quadratic function as shown in Fig.~\ref{fig1}b. 


\section{Mean-field hydrodynamics} \label{appendix_sec2}
Considering a broken symmetry homogeneous steady-state solution of the hydrodynamic EOMs, as provided in the main text, 
and solving for the modulus $P$ and $P_r$ of the order parameter fields, we obtain 
\begin{eqnarray} 
 \alpha_1\left(\bar\rho\right)P + \gamma_1\bar\rho P_r &&= \beta_1\left(P+P_r\right)^2P, \label{appeq1} \\
 \left(P+P_r\right)^2 &&= \frac{\gamma_2\rho_rP}{\beta_2P_r}. \label{appeq2}
\end{eqnarray} 
Plugging the expression for $\left(P+P_r\right)^2$ from Eq.~(\ref{appeq2}) into Eq.~(\ref{appeq1}), we obtain 
\begin{eqnarray}
 P_r = B P, \quad {\text {where }} B = \frac{\beta_1\gamma_2}{\alpha_1\left(\bar\rho\right)\beta_2} \rho_r. \label{appeq3}
\end{eqnarray}
Using Eq.~(\ref{appeq3}), we obtain the expression for $P^2$, as in Eq.~(\ref{eqPsquare}). This expression suggests 
enhance in the robustness of the system in the presence of the NRs.


\section{Linearised hydrodynamics} \label{appendix_sec3}
We consider fluctuations in the slow fields, as written in the main text. Incorporating these fluctuations in the EOM for 
${\bm P}_r$, and writing it for the $\parallel$ and $\perp$ components upto linear order in $B$, we obtain 
\begin{eqnarray}
 \delta P_{r \parallel} = A \delta P_{\parallel}, \quad {\text{where }} A = \frac{\gamma_2\rho_r-2\beta_2BP^2}{\beta_2(1+4B)P^2}, \label{appeq6}
\end{eqnarray}
and
\begin{eqnarray}
 \delta P_{r \perp} = B \delta P_{\perp}. \label{appeq7}
\end{eqnarray}
Similarly, writing the EOM for ${\bm P}$ for the $\parallel$-components and neglecting the higher order terms in fluctuations and derivatives, we get 
\begin{eqnarray}
 \delta P_{\parallel} = \frac{1}{\alpha_1^{\prime}-A^{\prime}}\left(\gamma_1BP-\frac{v_s}{2\bar\rho}\partial_{\parallel}\right)\delta\rho, 
 \label{appeq7}
\end{eqnarray}
where
\begin{eqnarray}
 \alpha_1^{\prime} &&= -\alpha_1(\bar\rho) + \beta_1(1+B)^2P^2 + 2\beta_1(1+B)P^2, \quad \label{appeq8} \\
 A^{\prime} &&= \left\{ \gamma_1\bar\rho-2\beta_1(1+B)P^2 \right\} A. \label{appeq9}
\end{eqnarray}
Note that, $\alpha_1^\prime=2\alpha_1(\bar\rho)$ and $A^\prime$ vanishes for the clean system, and therefore, Eq.~(\ref{appeq7}) 
takes the familiar form as in ref.~\cite{tonertu1998}.

Now as we have obtained the expressions for the fluctuations $\delta P_\parallel, \delta P_{r\parallel}$ and 
$\delta P_{r\perp}$, we solve for fluctuations $\delta P_\perp$ and $\delta\rho$. Plugging the above expressions into 
the $\rho$-equation~(\ref{EOM1}), we obtain
\begin{eqnarray}
 \partial_t\delta\rho = \left( D_{\parallel}\partial_{\parallel}^2 + D_{\rho}\partial_{\perp}^2 \right)\delta\rho - 
                        Xv_sP\partial_{\parallel}\delta\rho - v_s\bar\rho\partial_{\perp}\delta P_{\perp}. \label{appeq10} \notag \\
\end{eqnarray}
Also, writing the ${\bm P}$-equation~(\ref{EOM2}) for the $\perp$-components upto linear order terms in fluctuations and simplifying it, we obtain 
\begin{eqnarray}
 \partial_t\delta P_{\perp} = D\nabla^2\delta P_{\perp} + \lambda_1P\partial_{\parallel}\delta P_{\perp} - 
                              \frac{v_s}{2\bar\rho}\partial_{\perp}\delta\rho. \qquad \label{appeq13} 
\end{eqnarray}
Following the customary, we write Eqs.~(\ref{appeq10}) and (\ref{appeq13}) in the Fourier space. Given a function 
$u({\bm r},t)$, its Fourier transform in space and time is defined as 
\begin{eqnarray}
 u({\bm q},\omega) = \int_{-\infty}^\infty dt d{\bm r} e^{i\omega t} e^{-i{\bm q}\cdot{\bm r}} u({\bm r},t).
\end{eqnarray}
Using this definition, we can write the EOMs for the fluctuations as 
\begin{eqnarray}
 {\bm M}{\bm \Phi} = {\mathcal O},     \label{appeq14}
\end{eqnarray}
where the fluctuation vector 
\begin{eqnarray}
{\bm \Phi}=\left( \begin{array}{cc} \delta\rho({\bm q},\omega) \\ \delta P_{\perp}({\bm q},\omega) \end{array} \right)
\end{eqnarray} 
and $\mathcal O$ represents the null vector. The coefficient matrix is given by 
\begin{eqnarray}
 &&{\bm M}({\bm q},\omega) = \notag \\ 
 \quad && \left[ 
 \begin{array}{cc}
  i\left(\omega-Xv_sPq_{\parallel}\right)-\Gamma_{\rho} & -iv_s\bar\rho q_{\perp} \\
  iv_sq_{\perp}/2\bar\rho & -i\left(\omega+\lambda_1Pq_{\parallel}\right) + \Gamma_P 
 \end{array} \right]. \qquad\quad \label{appeq15} 
\end{eqnarray}
We obtain the normal modes of Eq.~(\ref{appeq14}) by solving $Det\left[{\bm M}\right]=0$. 
This eventually gives a quadratic equation $\omega^2 + b\omega + c = 0$, where
\begin{eqnarray}
 b &&= \left(\lambda_1-Xv_s\right)Pq_\parallel + i\left(\Gamma_P+\Gamma_\rho\right), \notag\\ 
 c &&= -\lambda_1Xv_sP^2q_\parallel^2 - \frac{v_s^2}{2}q_\perp^2 -\Gamma_P\Gamma_\rho \notag \\ 
   && \qquad\qquad\quad +i\left(\lambda_1\Gamma_\rho-Xv_s\Gamma_P\right)Pq_\parallel. \notag
\end{eqnarray}
The solution of this quadratic equation gives two sound modes $\omega_\pm$ with the convection speeds $c_\pm$, 
as discussed in the main text.

\section{Fluctuations in the rotator update introduces randomness in the system} \label{appendix_sec4}
\begin{figure}[t]
\includegraphics[width=\linewidth]{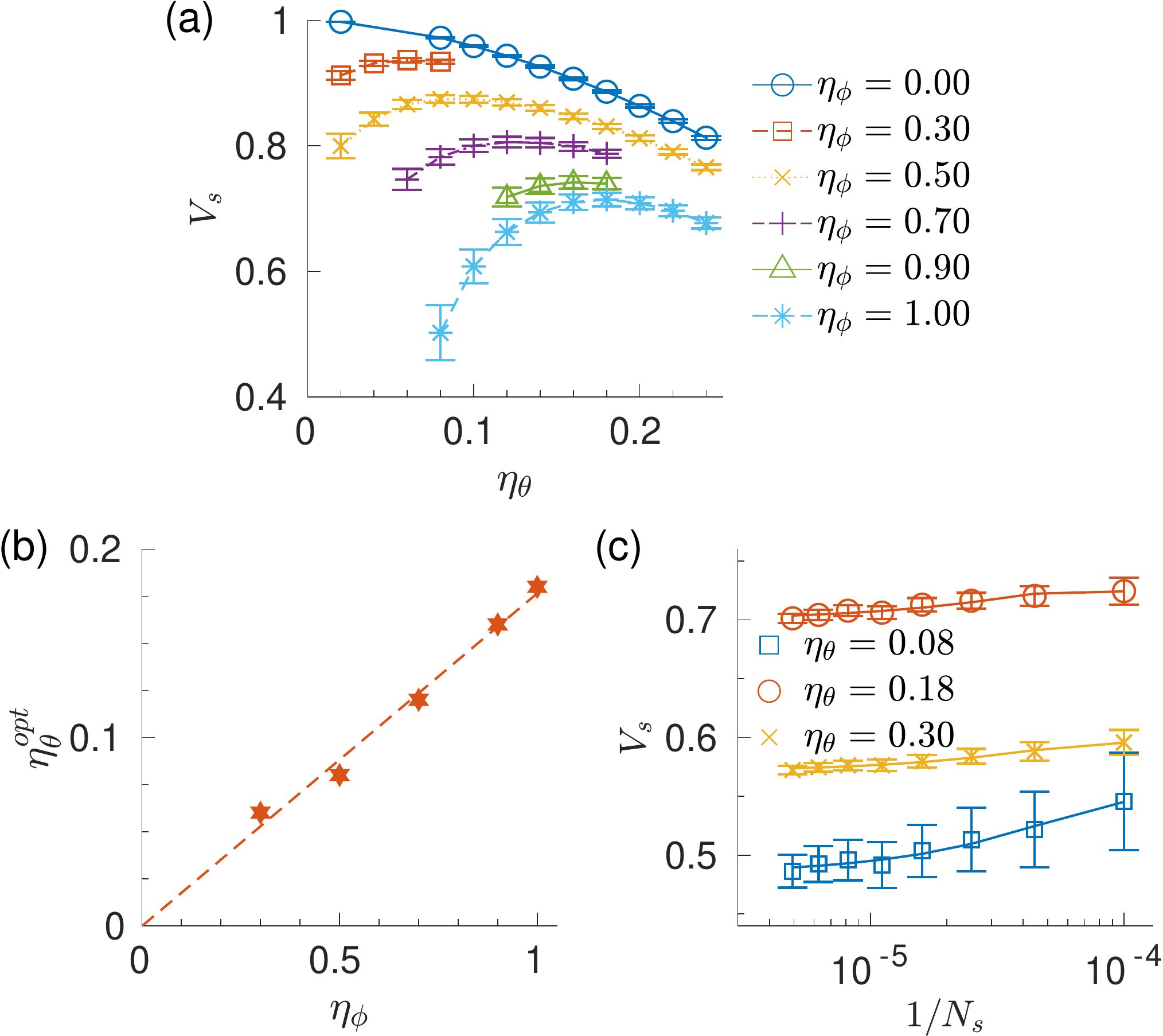}
\caption{
{$\eta_\phi$-term introduces fluctuations in the system.}
(a) $V_s$ versus $\eta_\theta$ plot is shown for $c_r=0.01, \alpha=1$, and various $\eta_\phi$. The SPPs obtain optimal 
ordering at a finite $\eta_\theta$ in the presence of the $\eta_\phi$-term. The curves are obtained near the optimal point 
only, and the maximum $V_s$ suggests respective $\eta_\theta^{opt}$.
(b) $\eta_\theta^{opt}$ varies linearly with $\eta_\phi$. The dashed line represents the extrapolation which  verifies 
$\eta_\theta^{opt}=0$ for $\eta_\phi=0$.
(c) Variation in $V_s$ with system size is shown on semi-log scale for $c_r=0.01, \alpha=1$ and $\eta_\phi=1$. 
The solid lines show fits quadratic in $1/N_s$ and have finite intercepts in the thermodynamic limit. 
}
\label{figsupple}
\end{figure}

Inclusion of the noise term $\eta_\phi\psi_\phi$ in the update Eq.~(\ref{updaterule3}) of NR-orientation induces the optimality feature, as the 
system attains the optimal ordering for a finite $\eta_\theta$ (Fig.~\ref{figsupple}a). The extra noise term in the $\phi$-update 
equation introduces randomness in the system that hinders mutual communications among the subflocks. Therefore, $V_s$ decreases 
with increasing $\eta_\phi$ (Fig.~\ref{figsupple}a). Provided the system has a finite $\eta_\theta$, the system overcomes 
the hindrance due to $\eta_\phi$ and attains the optimal order. The optimal noise $\eta_\theta^{opt}$ increases linearly with 
$\eta_\phi$ (Fig.~\ref{figsupple}b). An extrapolation of the linear fit ensures the monotonic order-disorder transition 
for $\eta_\phi=0$. 

Note that inclusion of the $\eta_\phi$-term 
makes $\phi$ behave like a colour noise to the SPP-orientation. As easily understood from the $\phi$-update rule, the 
autocorrelation of $\phi$ varies as $1/\eta_\phi^2$ for $\alpha=0$. This sets in a timescale in the system that inhibits 
us to obtain a true steady state in a reasonable cpu-time. However, for a finite $\alpha$, the NRs suppress the effect of 
the colour noise and we obtain steady-states. We show the variation in $V_s$ with system size  
for $\alpha=1$ and the maximum noise $\eta_\phi=1$ in Fig.~\ref{figsupple}c. We note that $V_s$ varies quadratically with 
$1/N_s$, and the corresponding fits give finite $V_s$ as $1/N_s=0$. Moreover, note that the fluctuation in $V_s$ decreases 
systematically with system size. Therefore, we argue that for finite $\alpha$ and finite $\eta_\phi$, the system will 
obtain a long-range order state; however, the presence of the $\eta_\phi$-term inhibits us to find it unambiguously.



\end{document}